# Eisenbasierte Vielfalt in der Supraleitung

Anna Böhmer und Andreas Kreyssig

Ames Laboratory, Ames, Iowa 50011, USA

Vor neun Jahren wurden auf dem magnetischen Element Eisen basierende Supraleiter entdeckt. Eine erstaunliche Vielzahl an chemischen Varianten und physikalischen Eigenschaften hält Wissenschaftler seitdem in Atem.

English summary:

## Unconventional superconductors: Iron-based diversity

Nine years ago, superconductors based on the magnetic element iron were discovered.  A flurry of research activity has revealed an unprecedented diversity of chemical structures and physical properties. Similarly to other unconventional superconductors, stripe-type antiferromagnetism seems to play an important role. Particularly interesting is its strong coupling to the crystal lattice. The systematic comparison of different compounds may allow to identify key elements in the mechanism of superconductivity and provide guidelines for the search for new and better superconductors.

Es scheint eine Regel zu geben, dass es im Feld der Supraleitung etwa alle 25 Jahre zu einer bedeutenden Entdeckung kommt. Am Anfang stand die Beobachtung des verschwindenden elektrischen Widerstands von Quecksilber bei etwa 4 K, also nur 4 Grad über dem absoluten Nullpunk, von Heike Kammerlingh-Onnes im Jahre 1911 [1]. Diese perfekte elektrische Leitfähigkeit liegt der Bezeichnung Supraleitung zugrunde. Eine zweite bedeutsame Entdeckung war, dass Supraleiter nicht nur Strom verlustfrei transportieren, sondern auch ein äußeres Magnetfeld verdrängen; dies ist der 1933 nachgewiesene Meissner Effekt. Er zeigt, dass Supraleitung eine thermodynamische Phase ist, genauso wie ‚flüssig' oder auch ‚ferromagnetisch'.

1957, knapp 50 Jahre nach der ursprünglichen Entdeckung, wurde mit der BCS Theorie (Bardeen, Cooper und Schrieffer) eine Erklärung dieses seltsamen Verhaltens gefunden. Supraleitung ist ein makroskopischer Quantenzustand, den man nur verstehen kann, wenn man den Wellencharakter der Elektronen und außerdem die Wechselwirkung zwischen Elektronen untereinander in Betracht zieht. An beides war 1911 noch nicht zu denken. Insbesondere die Beschreibung von wechselwirkenden Elektronen ist für Festkörpertheoretiker extrem anspruchsvoll. Sie sind daher auf die BCS Theorie besonders stolz. Es konnte gezeigt werden, dass eine – auch noch so kleine – anziehende Wechselwirkung zwischen Elektronen dazu führt, dass diese Paare formen. Diese sogenannten Cooperpaare kondensieren wiederum in einen geordneten Zustand (ähnlich wie kondensiertes Wasser mehr geordnet ist als Wasserdampf). Dies alles geschieht unterhalb einer kritischen Temperatur, der supraleitenden Sprungtemperatur. Das Kondensat kann Strom widerstandsfrei leiten und weist alle anderen mit Supraleitung verbundenen Quanteneigenschaften auf. Anhand verschiedener Möglichkeiten, wie sich die negativ geladenen Elektronen effektiv anziehen können, unterscheidet man konventionelle und unkonventionelle Supraleiter. In konventionellen Supraleitern gilt das berühmte Bild: ein Elektron passiert das Ionengitter des Metalls und deformiert es, indem es die positiv geladenen Gitterionen anzieht. Ist es vorbeigezogen, bleibt eine positive Ladungswolke, die wiederum ein zweites Elektron anzieht.

75 Jahre nach Kammerlingh-Onnes Entdeckung, wurde 1986 Hochtemperatursupraleitung in Kupraten gefunden, eine echte Überraschung. Plötzlich waren Übergangstemperaturen nicht mehr wenige Grad über absolut Null, sondern erreichten mehr als 100 K (Abbildung 1), eine extra Größenordnung! Zu verstehen, wie das möglich ist, ist auch heute noch eine Herausforderung. Kuprate sind ein Beispiel für unkonventionelle Supraleiter. Anders als in konventionellen Supraleitern, beruht hier die effektive Anziehung zwischen Elektronen nicht auf ihrer Wechselwirkung mit dem Ionengitter, sondern auf spezielle Wechselwirkungen zwischen den Elektronen selbst. [2].

Lange Zeit war ‚Hochtemperatursupraleiter' synonym zu ‚Kupratsupraleiter'. Das sollte sich aber 2008 ändern.

Eisenzeit der Supraleitung

Mit der Beobachtung von Supraleitung mit einer Sprungtemperatur von 26 K im Material La(O,F)FeAs [3] begann, was bald als ‚Eisenzeit der Supraleitung' bezeichnet wurde. Die überraschende Entdeckung ist auch ein Lehrstück dafür, die Augen offen zu halten. Eigentlich wurden nämlich magnetische Halbleiter gesucht und Hochtemperatursupraleitung in einem so eisenreichen Material war alles andere als erwartet. Eisen ist immerhin das bekannteste ferromagnetische Element und normalerweise sind

Ferromagnetismus und Supraleitung nicht verträglich miteinander: In einem klassischen Supraleiter bricht ein Magnetfeld, so wie es ein Ferromagnet erzeugt, die Cooperpaare auf und zerstört die Supraleitung.

Die supraleitende Sprungtemperatur von 26 K in La(O,F)FeAs ist bereits relativ hoch. Es begannen aber sofort Versuche die Sprungtemperatur weiter zu erhöhen. Ein typischer Ansatz ist zum Beispiel, mit extremem äußeren Druck die Supraleitung zu beeinflussen. Und tatsächlich: die Sprungtemperatur von La(O,F)FeAs steigt unter einem Druck von 4 GPa (dies entspricht dem Gewicht eines Airbus A380 auf einer Fläche von 10 cm$^2$) auf 43 K. Unter Druck wird das Kristallgitter komprimiert. Anstatt Druck von außen auszuüben, kann man einen ähnlichen Effekt daher auch erreichen, indem man ein Element durch ein anderes mit kleinerem Ionenradius ersetzt. Dies wurde auch sofort versucht und es konnte eine Rekord-Sprungtemperatur von 55 K erreicht werden, indem Lanthan (La) durch das kleinere Gadolinium oder Samarium ersetzt wurde [4].

Damit waren Kuprate als Hochtemperatursupraleiter endgültig nicht mehr allein. Da die Supraleitung in Kupraten immer noch nicht vollkommen verstanden war (und ist), kam nun die Hoffnung auf, durch den Vergleich mit einer zweiten Materialklasse die wesentlichen Voraussetzungen für Hochtemperatursupraleitung herausfiltern zu können. Es sollte aber so kommen, dass die eisenbasierten Materialien auch schon für sich genommen eine Vielzahl an faszinierenden Eigenschaften aufzeigen.

Materialienvielfalt

Der Reichtum an chemischen Varianten in der Materialklasse der eisenbasierten Supraleiter ist enorm (Abbildung 2). Schnell wurde bekannt, dass Supraleitung in LaOFeAs auf verschiedene Arten induziert werden kann. Eine Möglichkeit ist, wie in der Entdeckungsarbeit, einen kleinen Teil des Sauerstoffs durch Flour zu ersetzen. Das einwertige Flourion zieht ein Elektron weniger an sich als das zweiwertige Sauerstoffion. Damit stehen dem Leitungsband etwas mehr Elektronen zur Verfügung, ein Prozess, der dem Elektron-Dotieren von Halbleitern entspricht. Der gleiche Effekt wird auch durch Fehlstellen von Sauerstoff erzielt. Eine andere Möglichkeit ist aber auch, Lanthan teilweise durch Strontium zu ersetzen. In diesem Fall werden dem Leitungsband Elektronen entzogen und man spricht von Loch-Dotierung. Undotiertes LaOFeAs wird als „Mutterverbindung" bezeichnet, von denen die dotierten Systeme abgeleitet sind.

Das entscheidende strukturelle Element der eisenbasierten Supraleiter sind die Eisen-Arsen (FeAs) Lagen. Eisenatome befinden sich in einem ebenen quadratischen Gitter und Arsenatome sind tetraederartig um sie herum angeordnet. Die Zwischenlagen aus den anderen Atomen dienen dazu, das Ladungsverhältnis richtig einzustellen. Zum Beispiel ist LaO so eine Zwischenlage, die Ladungen an FeAs abgibt. Mit diesen Erkenntnissen konnte die Suche nach weiteren eisenbasierten Mutterverbindungen beginnen.

Die zweite Mutterverbindung, BaFe$_2$As$_2$, wurde bereits gezielt gefunden [5]. Die FeAs-Lagen dieser Verbindung sind nahezu identisch denen in LaOFeAs. Eine Abschätzung der Ladungsverteilung ergibt außerdem, dass Barium (Ba) die Rolle der LaO Lagen übernehmen kann. BaFe$_2$As$_2$ war 2008 zwar kristallographisch bekannt, die Eigenschaften bei tiefen Temperaturen waren aber praktisch nicht untersucht. Zunächst wurde gezeigt, dass BaFe$_2$As$_2$ ganz ähnliche physikalische Eigenschaften wie undotiertes LaFeAsO hat, ein sehr vielversprechendes Zeichen. Und nur wenige Tage später wurde bekannt, dass Lochdotierung (mittels des Ersetzens von Barium durch Kalium) tatsächlich Supraleitung in BaFe$_2$As$_2$ erzeugt, mit einer Sprungtemperatur von 38 K. Dass in dieser neuen Verbindung kein Sauerstoff enthalten ist, macht das Leben besonders für Materialpräparatoren deutlich einfacher. Aus den neuen Verbindungen lassen sich leicht große Kristalle herstellen, weshalb sie heute zu den am besten untersuchten Supraleitern zählen. BaFe$_2$As$_2$ beweist auch, dass für Hochtemperatursupraleitung Sauerstoff nicht prinzipiell nötig ist. Dies steht im Gegensatz zu Kupratsupraleitern, die allesamt Oxide sind.

Kurz darauf oder parallel wurden weitere Mutterverbindungen entdeckt. Sehr ähnlich zu BaFe$_2$As$_2$ sind zum Beispiel SrFe$_2$As$_2$ und CaFe$_2$As$_2$. Aber auch Systeme mit anderen Zusammensetzung wurden gefunden, wie LiFeAs, oder auch Verbindungen mit sehr komplizierten Zwischenlagen, wie zum Beispiel Sr$_3$Sc$_2$O$_5$Fe$_2$As$_2$ oder Ca$_{10}$(Pt$_3$As$_8$)[Fe$_2$As$_2$]$_5$. Eine Besonderheit ist das Material FeSe, welches überhaupt keine Zwischenlagen besitzt und in dem Selen (Se) den Arsen (As) -Platz besetzt. Die richtige Ladungsmenge in den Eisen-Ebenen wird durch den Unterschied in der Wertigkeit zwischen As und Se erreicht, die nebeneinander im Periodensystem stehen. Mit all diesen Materialien wurde offensichtlich eine Klassifikation von Mutterverbindungen nötig. Man unterscheidet sie anhand ihrer Stoichiometrie. In LaOFeAs stehen die vier Elemente im Verhältnis 1:1:1:1. Derartige Verbindungen werden daher als 1111-Systeme bezeichnet. BaFe$_2$As$_2$ ist dementsprechend ein 122-System, LiFeAs ein 111-System, FeSe ein 11-System und so weiter [6].

Ein bisschen scheint es als ob, egal was man tut, diese Mutterverbindungen immer supraleitend werden können. Eine ganze Reihe von Dotierungen sind möglich: BaFe$_2$As$_2$ ist ein besonders reichhaltiges Beispiel, sowohl Kalium oder Natrium für Barium führen zum Ziel (Lochdotierung), als auch Cobalt, Nickel, Rhodium, Palladium oder Iridium für Eisen (Elektrondotierung). Auch partielle Substitution von Elementen in der gleichen Spalte des Periodensystems, was die Ladungsverteilung nicht offensichtlich verändert, kann Supraleitung induzieren, z.B. das Ersetzen von Eisen durch Ruthen oder von Arsen durch Phosphor. Neben den diversen Arten von Dotierung funktioniert auch wieder extremer äußerer Druck zum Induzieren von Supraleitung.

Überraschenderweise können sogar Bier, Wein, Whisky, Sake und Shochu (letztere sind japanischer Reiswein bzw. Reisschnapps) Supraleitung induzieren, und zwar in dem mit FeSe verwandten Material Fe(Te$_{0.8}$S$_{0.2}$) [7]. Am liebsten „mag" der Supraleiter anscheinend Rotwein, was zur höchsten Sprungtemperatur führt, und am wenigsten Shochu. Alle Getränke funktionieren besser als das entsprechende Wasser-Ethanol Gemisch. Der Effekt wurde durch eine gründliche Nachfolgearbeit entmystifiziert. Das Ausgangsmaterial Fe(Te$_{0.8}$S$_{0.2}$) hat nämlich das Problem, dass sich einige Eisenatome zuviel im Material befinden. Diese sitzen auf Zwischenplätzen im Gitter und stören die Supraleitung. Bestimmte organische Säuren, die insbesondere in Rotwein vorkommen entziehen sie dem Material, und damit kann sich Supraleitung ungestört ausbilden.

Damit war innerhalb weniger Monate ein ganzer Zoo eisenbasierter Supraleiter bekannt. Was auf der größeren Ebene erhofft war, nämlich aus dem Vergleich zwischen Kupraten und eisenbasierten Materialien zu lernen, begann nun schon im Kleinen. Die verschiedenen Mutterverbindungen, die verschiedenen Dotierungen, alle können auf der Suche nach allgemeingültigen Prinzipien miteinander verglichen werden. Eine Theorie, die ein Material beschreibt, kann an allen anderen Materialien getestet werden. Und es braucht nur ein Gegenbeispiel, um die Allgemeingültigkeit einer Theorie zu widerlegen.

Eine Frage ist zum Beispiel, ob es einen strukturellen Parameter gibt, der die supraleitende Sprungtemperatur optimiert. Verschiedene Kandidaten wurden diskutiert. Zum Beispiel gab es die Idee, dass die Sprungtemperatur am höchsten ist, wenn die Eisen-Arsen-Tetraeder möglichst regelmäßig sind, also alle Arsen-Eisen-Arsen Winkel den Wert des idealen Tetraederwinkel von 109,5° haben [8]. Dies ist in dem Kalium-dotierten $BaFe_2As_2$ System und in vielen 1111 Systemen tatsächlich gegeben. Es gibt aber auch klare Gegenbeispiele, zum Beispiel $BaFe_2(As,P)_2$ oder $Ba(Fe,Co)_2As_2$. Damit ist diese Theorie schon nicht mehr allgemein gültig.

Vielfältige Phasen

Die meisten bekannten Metalle zeigen beim Abkühlen keine Änderung ihres Zustands. Die Eigenschaften eines Kupferkabels unterscheiden sich nicht grundsätzlich, ob bei Raumtemperatur oder bei Temperaturen nahe des absoluten Nullpunkts. Ein Beispiel für ein Material, dessen Eigenschaften sich bei Temperturänderung grundlegend ändern, ist elementares Eisen. Eisen ist ein ‚normales Metall' nur oberhalb von 770°C, darunter ordnet es ferromagnetisch. Viele, insbesondere konventionelle Supraleiter, sind oberhalb ihrer Sprungtemperatur normale Metalle. Festkörperphysiker spitzen aber die Ohren, wenn Materialien magnetisch *und* supraleitend sind, beziehungsweise, wenn sie magnetisch geordnet sind und nur durch eine kleine Änderung, z.B. der Dotierung, supraleitend werden. Hier besteht die Möglichkeit, einen unkonventionellen Supraleiter gefunden zu haben.

Wie schon erwähnt, ist mit Supraleitung verträgliche Magnetismus nur in den seltensten Fällen Ferromagnetismus. Dass liegt daran, dass im Ferromagneten alle Elektronenspins (die „magnetischen Kompassnadeln" der Elektronen) parallel ausgerichtet sind, in Cooperpaaren müssen sie aber antiparallel sein (außer in den exotischsten Fällen). Anders als im Ferromagnetismus, können in einem antiferromagnetischen Material jeweils die Hälfte aller Spins entgegengesetzt ausgerichtet sein. Abbildung 3 (unten) zeigt ein Beispiel für eine antiferromagnetische Spinordnung, so wie sie in vielen eisenbasierten Materialien beobachtet wird. In diesem Fall steht der Bildung von Cooperpaaren prinzipiell nichts im Wege. Viele Theorien gehen sogar davon aus, dass Antiferromagnetismus die Ursache für unkonventionelle Supraleitung sein kann. Genauer gesagt braucht es dafür einen „fast antiferromagnetischen" Zustand, das ist ein Zustand der kurz vor antiferromagnetischer Ordnung steht. In diesem Fall treten Fluktuationen (kurzzeitig und auf kleinem Raum antiferromagnetisch geordnete Bereiche) auf. Diese Fluktuationen können die Rolle der kurzzeitigen Gitterdeformationen des konventionellen Supraleiters übernehmen und Cooperpaarbildung vermitteln.

Vieles deutet darauf hin, dass die Supraleitung in eisenbasierten Materialien tatsächlich durch solche antiferromagnetischen Fluktuationen vermittelt wird. Die meisten Mutterverbindungen sind nämlich unterhalb von etwa 100 K – 200 K antiferromagnetisch geordnet. Äußerer Druck oder Dotierung verringert diese Temperatur, die Ordnung wird unterdrückt. Wenn dies ausreichend geschehen ist, tritt die Supraleitung ein. Dies ist typisch für einen unkonventionellen Supraleiter [2].

Solche magnetischen, und auch die allgemeiner unkonventionellen, Supraleiter haben deshalb interessante und oft ähnliche Phasendiagramme. Mit Phase bezeichnet man einen bestimmten Zustand eines Materials, zum Beispiel sind ‚Eis' und ‚Wasser' die festen und flüssigen Phasen von $H_2O$. Phasendiagramme, wie in den Abbildungen 4, 5 und 6, sind dann eine Art Landkarten, die zeigen, welcher Zustand bei einer bestimmten Temperatur und einem bestimmten Wert eines äußeren Stellparameters stabil ist. Beispiele für solche Stellparameter sind die schon beschriebene Dotierung und äußerer Druck. In dem typischen Phasendiagramm eines unkonventionellen Supraleiters grenzt Supraleitung fast immer an eine antiferromagnetische Phase. Dem gerade beschriebenen Bild entsprechend ist Supraleitung oft am stärksten und die Sprungtemperatur am höchsten, wenn die antiferromagnetische Ordnung gerade unterdrückt ist und damit ihre Fluktuationen am stärksten sind. Bei weiterer Vergrößerung des Stellparameters wird die Sprungtemperatur wieder kleiner.

In vielen eisenbasierten Materialien gibt es einen spannenden Koexistenzbereich von Supraleitung und antiferromagnetischer Ordnung. So eine Koexistenz tritt nicht notwendigerweise auf. Wenn es sie aber gibt, lassen sich interessante Effekte im Phasenwechselspiel beobachten. Faszinierend ist zum Beispiel, dass das geordnete magnetische Moment von der Supraleitung deutlich verringert wird. Dies bedeutet, dass die Supraleitung die magnetische Ordnung schwächt. Man sagt, die beiden Phasen stehen im Wettbewerb [9].

Kopplung von Magnetismus und Kristallgitter

Die Form der antiferromagnetischen Ordnung in eisenbasierten Supraleitern hat ebenfalls spannende Konsequenzen. Im Allgemeinen können Antiferromagneten verschiedene „Muster" haben, solange nur die Vektorsumme aller Spins Null ergibt. In eisenbasierten Materialien gibt es da eine sogenannte Streifenordnung. Entlang einer Richtung im Kristall zeigen alle Spins in die gleiche Richtung, entlang der anderen Richtung alternieren sie (Abbildung 3 unten). Damit unterscheidet sich natürlich die eine Richtung von der anderen. Eisenatome mit parallelen Spins haben eine andere Beziehung zueinander als Eisenatome mit antiparallelen Spins, und als Konsequenz haben sie jeweils einen anderen Abstand zueinander. Damit ist das Gitter der Eisenatome nicht mehr quadratisch, sondern leicht rechteckig. Man sagt, die quadratische Symmetrie des atomaren Gitters ist gebrochen. Das bedeutet, dass mit dem antiferromagnetischen Ordnung auch eine Gitterverzerrung verbunden ist.

Große Aufmerksamkeit haben Variationen der gerade geschriebenen magneto-strukturellen Kopplung erfahren. In einigen Systemen, wie zum Beispiel Cobalt-dotiertes $BaFe_2As_2$, gibt es nämlich noch einen Zwischenzustand (Abbildung 3, Mitte). Beim Abkühlen tritt zuerst die Gitterverzerrung auf und erst typischerweise 5 – 10 K später die magnetische Streifenordnung. Diese Beobachtung lässt Zweifel aufkommen, ob die Gitterverzerrung tatsächlich nur eine Folge des streifenartigen Magnetismus ist oder ob sie eine andere Ursache hat.

Allerdings muss die Idee, dass der streifenartige Magnetismus hinter der Gitterverzerrung steht, nicht aufgegeben werden. Es konnte nämlich theoretisch gezeigt werden [10], dass allein die Tendenz, magnetische Streifen zu bilden, ausreicht, eine Gitterverzerrung zu induzieren. Diese streifenartigen antiferromagnetischen Fluktuationen, bei denen sich die Spins nur kurzzeitig und in einem begrenzten Bereich in Streifen anordnen, können auch die quadratische Symmetrie brechen, selbst wenn die Spins noch nicht in Streifen angeordnet sind. Eine ähnliche Situation gibt es in Flüssigkristallen. Flüssigkristalle bestehen aus länglichen Molekülen. Diese zeigen neben eines komplett ungeordneten Zustands (Flüssigkeit) und eines komplett geordneten Zustands (Kristall) auch partiell geordnete Zustände. Zum Beispiel können alle Moleküle in eine bestimmte Richtung ausgerichtet sein, ohne dass sie aber feste Plätze wie in einem Kristallgitter einnehmen. Damit ist eine Richtung anders als die anderen Richtungen, die Symmetrie ist gebrochen. Diese Phase heißt nematisch. In Anlehnung an diese Flüssigkristallphase wird auch die rechteckig verzerrte Phase der eisenbasierten Supraleitern nematisch genannt.

Die Idee, dass ungeordnete, fluktuierende magnetische Momente eine Gitterverzerrung hervorrufen können, ist schon an sich sehr interessant. Der Versuch, diese Theorie zu belegen bzw. zu widerlegen, hat viel Kreativität angeregt. Bislang sind die allermeisten experimentellen Ergebnisse mit ihr vereinbar [9], aber wie schon erwähnt, braucht es nur ein Gegenbeispiel, um einer Theorie ihre Allgemeingültigkeit zu nehmen.

Ein Kandidat für ein solches Gegenbeispiel, nämlich dass es Nematizität ohne magnetische Fluktuationen geben kann, ist das Material FeSe. In FeSe tritt die rechteckige nematische Gitterverzerrung ganz klar unterhalb von 90 K auf. Aber selbst, wenn es bis zu den tiefsten Temperaturen abgekühlt wird, bildet sich keine magnetische Ordnung aus. Dies unterstützt die Idee, dass die Gitterverzerrung unhängig vom Antiferromagnetismus auftreten kann. Nun kann man aber versuchen nachzuweisen, ob Antiferromagnetismus trotzdem ‚in der Nähe' ist. So kann man testen, ob sich mithilfe eines Stellparameters magnetische Ordnung induzieren lässt, z.B. mittels äußeren Drucks. Und tatsächlich bildet sich unter Druck magnetische Ordnung in FeSe aus. Das Phasendiagram von FeSe unter Druck ist dem Phasendiagram von Cobalt-dotiertem $BaFe_2As_2$, zumindest eingeschränkt, ähnlich (Abbildung 5). Im Detail bleibt aber die Frage offen, ob die magnetische Ordnung in FeSe derselbe Streifen-Antiferromagnetismus ist, der auch in den anderen Systemen auftritt. Wir haben kürzlich zeigen können, dass die magnetische Ordnung unter Druck in FeSe den gleichen Effekt auf das Gitter der Eisenatome hat wie in den anderen Materialien [11]. Tritt sie auf, wird das quadratische Gitter sprunghaft rechteckig verzerrt. Ist das Gitter bereits rechteckig, vergrößert sie diese Verzerrung. Damit erscheint es möglich, dass sich das allgemeinere Prinzip wieder herstellen lässt, und die Gitterverzerrung in eisenbasierten Supraleitern immer an streifenartige antiferromagnetische Fluktuationen gekoppelt ist.

Es gibt noch eine andere Variante der magnetischen Ordnung in eisenbasierten Supraleitern, die sich spektakulär in den Atomabständen widerspiegelt. Wie ausführlich beschrieben, ist mit dem Streifenmuster der magnetischen Momente der Eisenatome eine Gitterverzerrung verbunden. Dabei wählt das Material eine der beiden möglichen, senkrecht zueinander stehenden, Richtungen der Streifen ‚spontan' aus. Lochdotierte eisenbasierte Supraleiter vom 122-Typ scheinen sich manchmal aber nicht entscheiden zu können und wählen stattdessen *beide* Richtungen gleichzeitig. Es ergibt sich damit eine quantenmechanische Überlagerung von zwei magnetischen Streifenstrukturen. Damit kann das Eisengitter seine quadratische Symmetrie beibehalten. Dieser Zustand wird ‚C4' Phase genannt (Abbildung 6). Spannend ist, dass dieser quadratische Zustand nur bei bestimmten Temperaturen auftritt, bei anderen Temperaturen gibt es die normale ‚einstreifige' Ordnung. Beim Abkühlen erscheint diese zuerst und das Gitter wird rechteckig, aber dann kommt die zweite Sorte Streifen dazu, und das Gitter wird wieder quadratisch. Und, um dem Phasen-Wechselspiel die Krone aufzusetzen, tritt dann bei noch tieferen Temperaturen auch noch Supraleitung auf. Anscheinend bevorzugt Supraleitung klar die nur einstreifige Ordnung, da diese Form unterhalb der supraleitenden Sprungtemperatur wieder erscheint und das Gitter gleichzeitig wieder rechteckig wird [12]. Dieses vielfältige Wechselspiel kann man zum Beispiel in der Probenlänge ganz deutlich sehen (Abbildung 6).

Probenlänge als Messgröße

Eine besonders anschauliche und aussagekräftige Messgröße ist die Probenlänge und deren thermische Ausdehnung. Die meisten Materialien dehnen sich beim Aufwärmen aus, z.B. kann extreme Hitze Eisenbahnschienen verformen. Auch bei einem Phasenübergang kann sich die Probenlänge plötzlich ändern. Man denke nur an gefrierendes Wasser, welches durch seine Ausdehnung Wasserleitungen zum

Bersten bringt. So gibt es bei fast jedem Phasenübergang Längenänderungen. Manchmal sind diese Längenänderungen nur sehr klein, in der Größenordnung von einem tausendstel Prozent und weniger. Sie können aber trotzdem mit spezialisierten Instrumenten gemessen werden. Eine hochsensitive Messmethode nutzt zum Beispiel den wohlbekannten, simplen Plattenkondensator. Die Kapazität eines Plattenkondensators ist umgekehrt proportional zum Plattenabstand und Kapazitäten lassen sich sehr präzise messen. Jetzt muss nur noch mechanisch die Probenlänge mit dem Plattenabstand gekoppelt werden um sie hochpräzise zu messen. Dies geschieht mithilfe einer geschickten Federführung (Abbildung 6, links). Grundsätzlich lässt sich das Prinzip aber auch in einfacheren Aufbauten ausnutzen.

Der antiferromagnetische Phasenübergang der eisenbasierten Supraleiter ändert die Abstände zwischen den Eisenatomen, was man als Längenänderung der gesamten Probe beobachten kann. Die Länge eines Kristalls ist nämlich nichts anderes als das Produkt aus dem durchschnittlichen Abstand zwischen den Atomen im Kristallgitter und der Anzahl dieser Atome. Abbildung 6 zeigt die Änderung der Probenlänge entlang der magnetischen Streifen in dem lochdotiertem 122-System $Ba_{1-x}K_xFe_2As_2$. Sobald die magnetischen Momente in Streifen ordnen, verkürzt sich diese Länge. Insbesondere ist der Wechel zwischen einstreifiger und zweistreifiger magnetischer Ordnung spektakulär sichtbar. Die durch die Streifenordnung verkürzte Länge wird beim Eintreten der zweistreifigen Ordnung (C4 Phase) länger und, wenn bei tieferen Temperaturen gleichzeitig mit Supraleitung die einstreifige Ordnung wieder einsetzt, wiederum kürzer [12].

Erstaunlich ist, dass auch das Einsetzen von Supraleitung praktisch immer in der Probenlänge zu sehen ist. Dies hat mit dem schon beschriebenen Einfluss von äußerem Druck auf die Sprungtemperatur zu tun. So setzt eine intuitive Beziehung die Längenänderungen bei einem Phasenübergang mit dessen Druckabhängigkeit in Verbindung: Wenn eine Phase ein kleineres Volumen hat, dann wird äußerer Druck (der das Volumen reduziert) diese Phase gegenüber der Phase mit größerem Volumen bevorzugen. Damit ändert äußerer Druck die Übergangstemperatur. Um wieder gefrierendes Wasser als Beispiel zu nehmen: Eis hat ein größeres Volumen als flüssiges Wasser. Äußerer Druck bevorzugt daher die flüssige Phase und bringt Eis zum Schmelzen. Wenn also äußerer Druck die supraleitende Sprungtemperatur erhöht, liegt das daran, dass die supraleitende Phase ein kleineres Volumen hat als die normalleitende Phase. In dem entgegengesetzten Fall, in dem die supraleitende Phase ein größeres Volumen hat als die normalleitende Phase, verringert sich die Sprungtemperatur unter Druck. Beide Sorten von Verhalten wurden in eisenbasierten Materialien beobachtet. In dem Beispiel in Abbildung 6 ist klar zu erkennen, dass die Probe im supraleitenden Zustand länger wird.

Dieses reichhaltige Wechselspiel zwischen Magnetismus, Gitterverzerrung und Supraleitung ist ein Grund für die Faszination von eisenbasierten Supraleitern. Trotz einer Vielzahl an Varianten ist selbst in den Ausnahmefällen Supraleitung immer in der Nähe von streifenartigem Antiferromagnetismus zu finden. Dies ist ein deutlicher Hinweis darauf, dass diese antiferromagnetischen Fluktuationen der Cooperpaarbildung und damit der Supraleitung zugrunde liegen.

## Ausblick

Kurz nach der Entdeckung der eisenbasierten Supraleiter kam die Hoffnung auf, dass sie sich auch für Anwendungen als nützlich erweisen werden. Viele Anwendungen von Supraleitung nutzen den verlustfreien Stromtransport um, wie in einem Elektromagneten, starke Magnetfelder zu erzeugen. Dies passiert zum Beispiel in einem Magnetresonanz-Tomographen im Krankenhaus. Supraleiter tragen auch die Ströme, die das Magnetfeld zu erzeugen, das die Elementarteilchen in Ringbeschleunigern wie im CERN auf der Bahn hält. Bis heute werden in den meisten Anwendungen klassische Supraleiter verwendet. Im allgemeinen gibt es, neben einer möglichst hohen Sprungtemperatur, noch zwei weitere Eigenschaften, die ein Supraleiter braucht, um für Anwendungen geeignet zu sein. Erstens, kann ein Supraleiter nur ein bestimmtes „oberes kritisches" magnetisches Feld ertragen, bevor die Supraleitung zusammenbricht. Damit ist dieses Feld das maximale, was eine supraleitenden Spule aus diesem Material erzeugen kann. Desweiteren können Supraleiter auch nur eine maximale „kritische" Stromstärke verkraften, bis die Supraleitung zerstört wird.

Die großen Vorteile von Kupratsupraleitern gegenüber klassischen Supraleitern sind ihre viel größeren Sprungtemperaturen und ihr viel höheres oberes kritisches Magnetfeld. Allerdings ist das obere kritische Feld von Kupratensupraleitern extrem abhängig von dem Winkel zwischen den Magnetfeldlinien und dem Kristallgitter. Dies stellt eine große technische Herausforderung für die Fertigung von Drähten dar. Zudem sind Kuprate Keramiken und daher spröde. Sie lassen sich deshalb nur schwer verarbeiten.

Im Vergleich zu Kupraten haben eisenbasierte Materialien klare Vorteile. Ihre Sprungtemperatur und ihre oberen kritischen Magnetfelder sind hoch, wenn auch geringer als in Kupraten. Allerdings sind die kritischen Felder der eisenbasierten Supraleiter deutlich weniger richtungsabhängig, was ihre Anwendung deutlich vereinfachen sollte. Als Metalle sind eisenbasierte Supraleiter zudem duktil (was mit ihrer großen Druckabhängigkeit zusammenhängt) und ihre Verarbeitung ist einfacher als die von Kupraten. Schließlich gilt es, eine ausreichende kritische Stromstärke zu erzielen, was mit Materialoptimierung erreicht werden kann. Diese Arbeit hat bei den Kupratsupraleitern Jahrzehnte gedauert. Während sie bei eisenbasierten Supraleitern noch läuft, haben vielversprechende Experimente bereits gezeigt, dass es prinzipiell möglich ist, technisch relevante Stromstärken auch mit eisenbasierten Supraleitern zu erreichen. Damit bieten sie das Potential für die Entwicklung neuer extrem starker Magnete [13].

Auf jeden Fall stellen die eisenbasierten Supraleiter ein großes Spielfeld für die Grundlagenforschung dar. Ihre vielfältigen und komplexen Eigenschaften haben neue Möglichkeiten zur systematischen Erforschung unkonventioneller Supraleitung eröffnet und die entwickelten Ideen bereichern die Studie auch anderer supraleitender Materialklassen. Inbesondere scheint die enge Verbindung zwischen Magnetismus und unkonventioneller Supraleitung fast universell zu sein. Es sieht daher tatsächlich so aus, dass magnetische Fluktuationen der Schlüssel zu Hochtemperatursupraleitung sind. Eine hohe Energie- bzw. die Temperaturskala dieser Fluktuationen scheint eine hohe supraleitende Sprungtemperatur zu ermöglichen. Bemerkenswert ist schließlich, dass als gemeinsames Motiv in der letzten Zeit sogar streifenartige Strukturen in anderen Materialklassen, insbesondere in Kupraten, gefunden wurden.

## Zusammenfassung

Supraleitung, obwohl schon über 100 Jahren bekannt, bleibt ein Feld voller Überraschungen. Seit neun Jahren stehen eisenbasierte Supraleiter im Zentrum eines atemberaubenden kollektiven Forschungsbemühen. Es erschloss eine unerwartete Vielfalt an chemischen Varianten und physikalischen Eigenschaften. Ähnlich wie bei anderen unkonventionellen Supraleitern scheint

Magnetismus, in diesem Fall mit einer ‚streifenartiger' Ausprägung und starker Wechselwirkung mit dem Kristallgitter, eine wichtige Rolle zu spielen. Der Vergleich verschiedener Systeme ermöglicht zunehmend zu klären, woher Supraleitung kommt und wie man sie steuern kann.

Abbildungen:

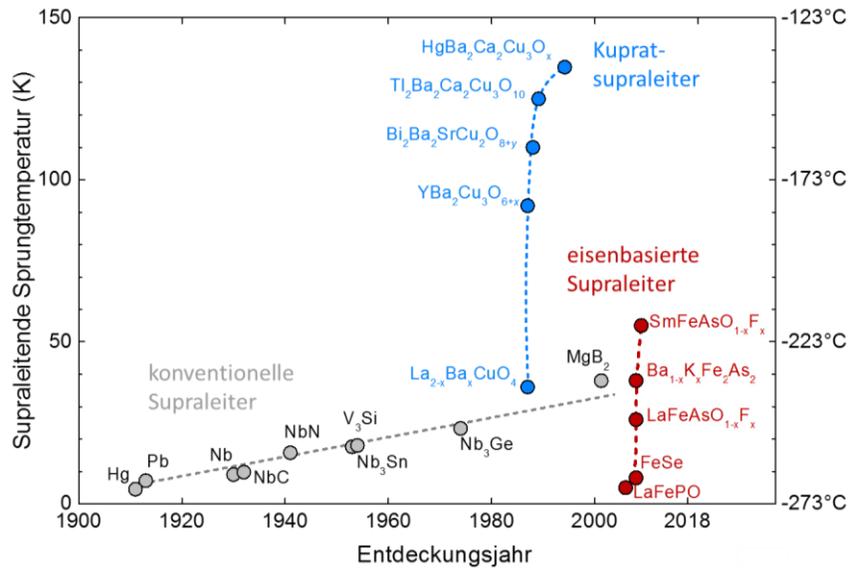

Abb. 1: Geschichte der Entdeckung ausgewählter supraleitender Materialfamilien.

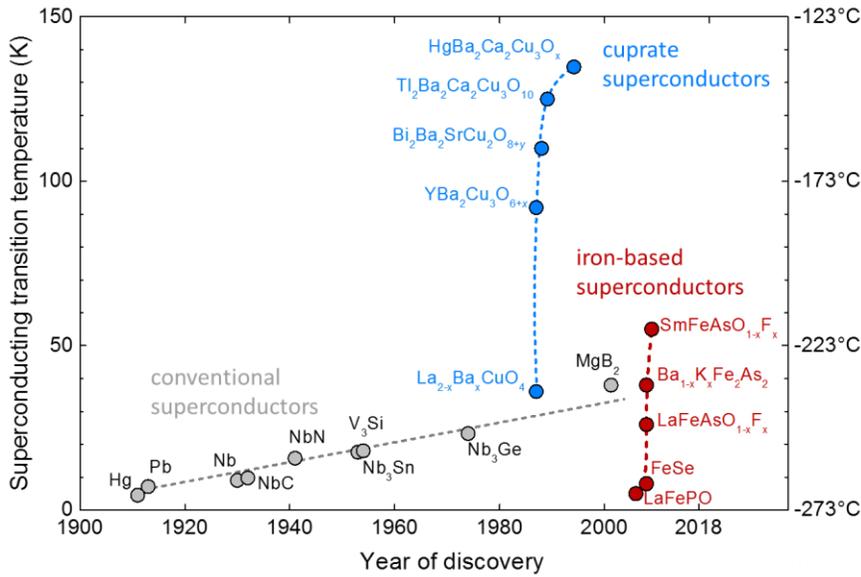

Fig. 1: History of the discovery of selected superconducting material families.

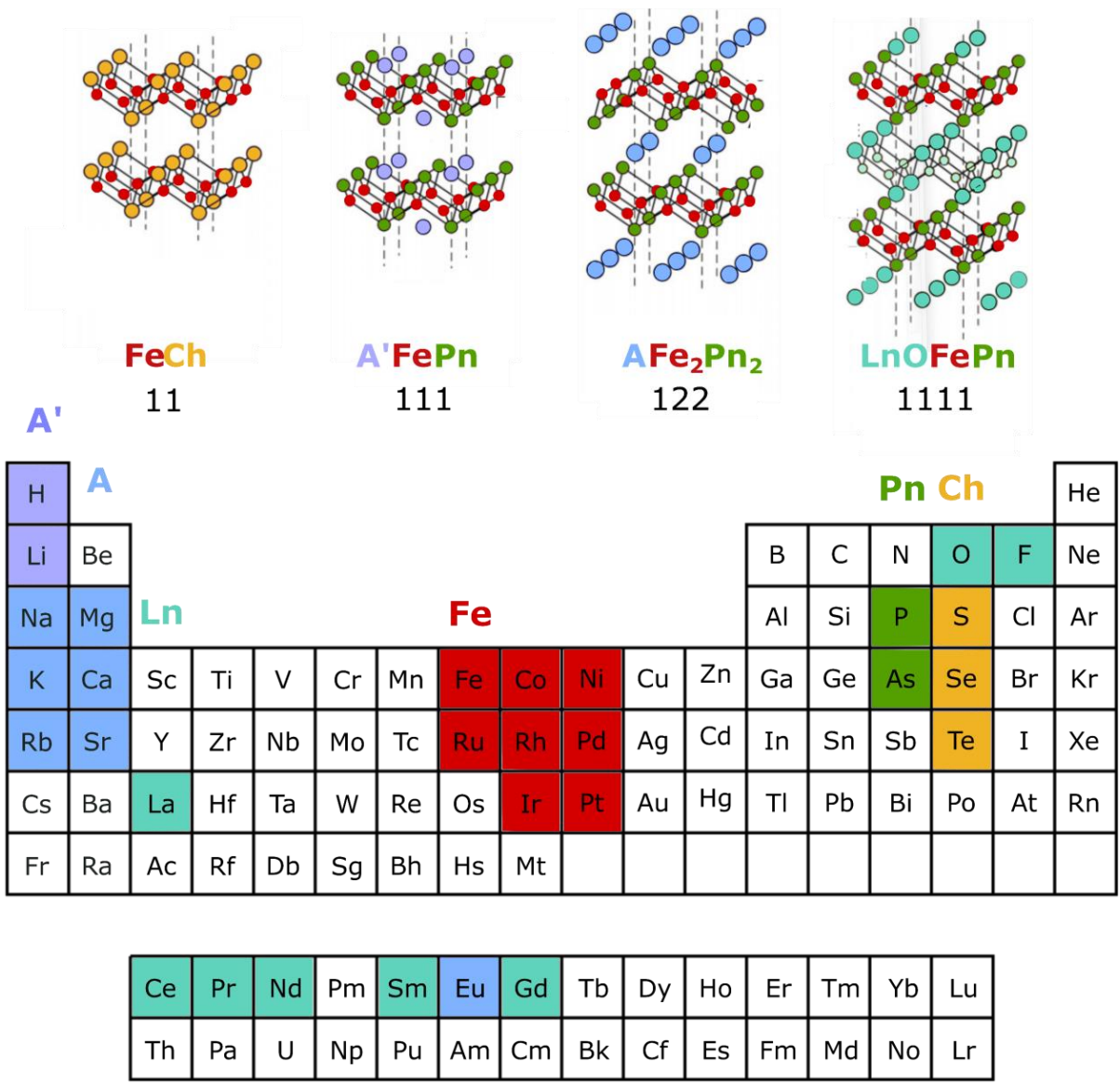

Abb. 2: Struktur und Elemente zum „Bau" eisenbasierter Supraleiter. Die atomare Gitterstruktur der vier bekanntesten Systeme ist oben dargestellt. Die eingefärbten Elemente des Periodensystems können die entsprechenden, identisch eingefärbten, Plätze des Atomgitter (partiell) besetzen.

Fig. 2: Structure and elements for the "construction" of iron-based superconductors. The atomic structure of the four most common systems is shown in the top row. The elements of the periodic table marked by a color can (partially) occupy the identically colored sites of the atomic lattice.

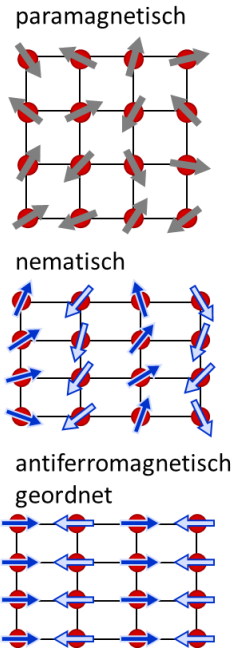

Abb. 3: Streifenartiger Antiferromagnetismus und rechteckige Verzerrung des atomaren Gitters der Eisenatome in eisenbasierten Materialien (unten). Der Hochtemperaturzustand ist ein quadratisches Gitter mit paramagnetischen, d.h. ungeordneten, Spins (oben). Im „nematischen" Zwischenzustand (Mitte) sind die Spins zwar ungeordnet, das Gitter aber trotzdem rechteckig.

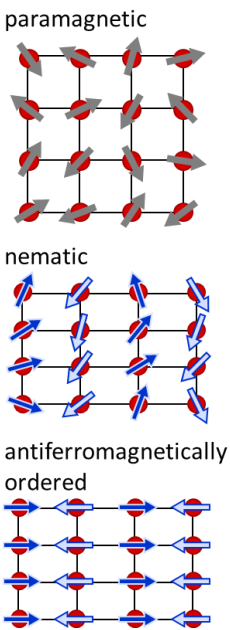

Fig. 3: Stripe-type antiferromagnetism and rectangular lattice distortion of the atomic iron lattice of iron-based materials (bottom). The high-temperature state is a square lattice with paramagnetic, i.e., disordered, spins (top). In a 'nematic' intermediate state (middle), the spins are disordered but the lattice is nevertheless rectangular.

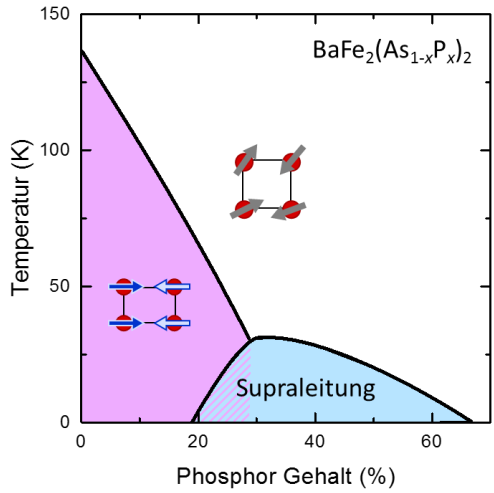

Abb. 4. Phasendiagram von Phosphor dotiertem BaFe$_2$As$_2$. Je nach Temperatur und Phosphor Gehalt ist das Eisengitter rechteckig/antiferomagnetisch (violetter Bereich), oder quadratisch/paramagnetisch.

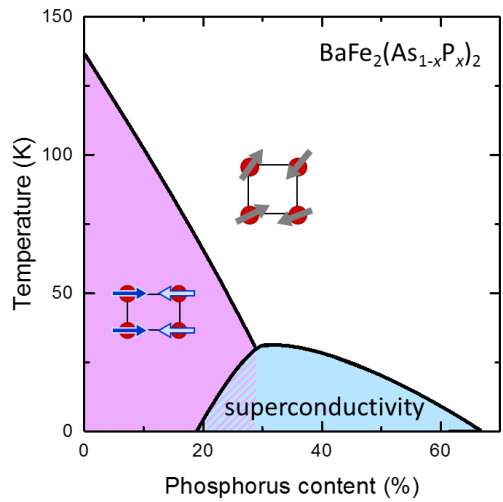

Fig. 4. Phase diagram of phosphorus-substituted BaFe$_2$As$_2$. Depending on temperature and phosphorus content, the iron lattice is either rectangular and antiferromagnetic (purple area), or quadratic and paramagnetic.

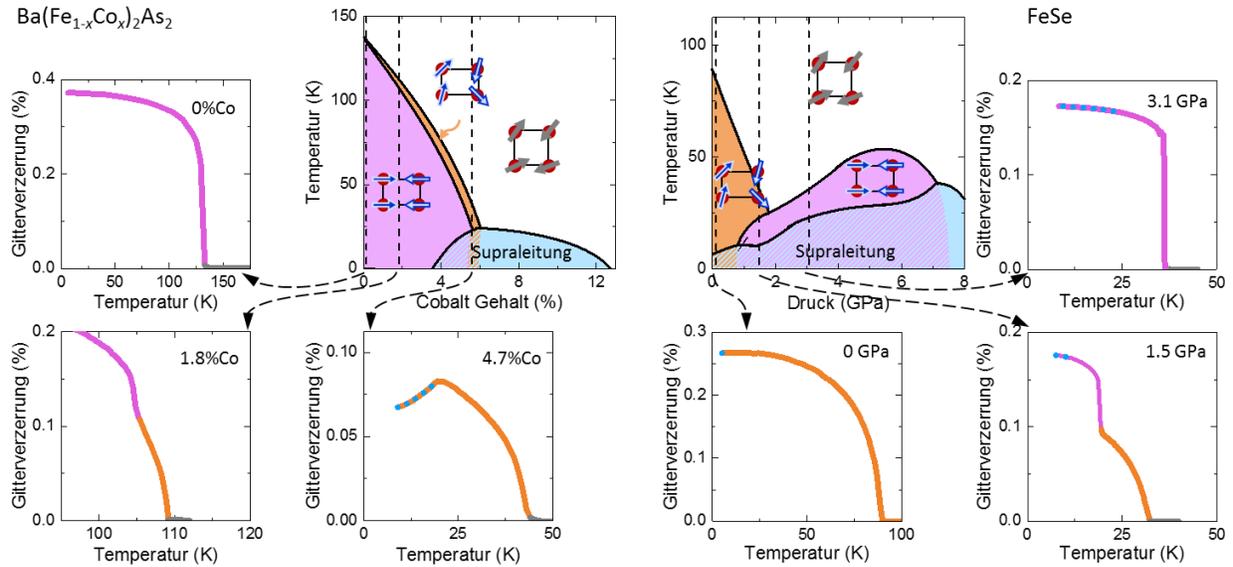

Abb. 5. Gitterverzerrung (relativer Längenunterschied der beiden Richtungen im rechteckigen Gitter) in Cobalt-dotiertem BaFe$_2$As$_2$ und in FeSe unter Druck, gemessen jeweils entlang der gestrichelten Linien im zugehörigen Phasendiagramm. Die Farben entsprechenden den entsprechenden Phasen.

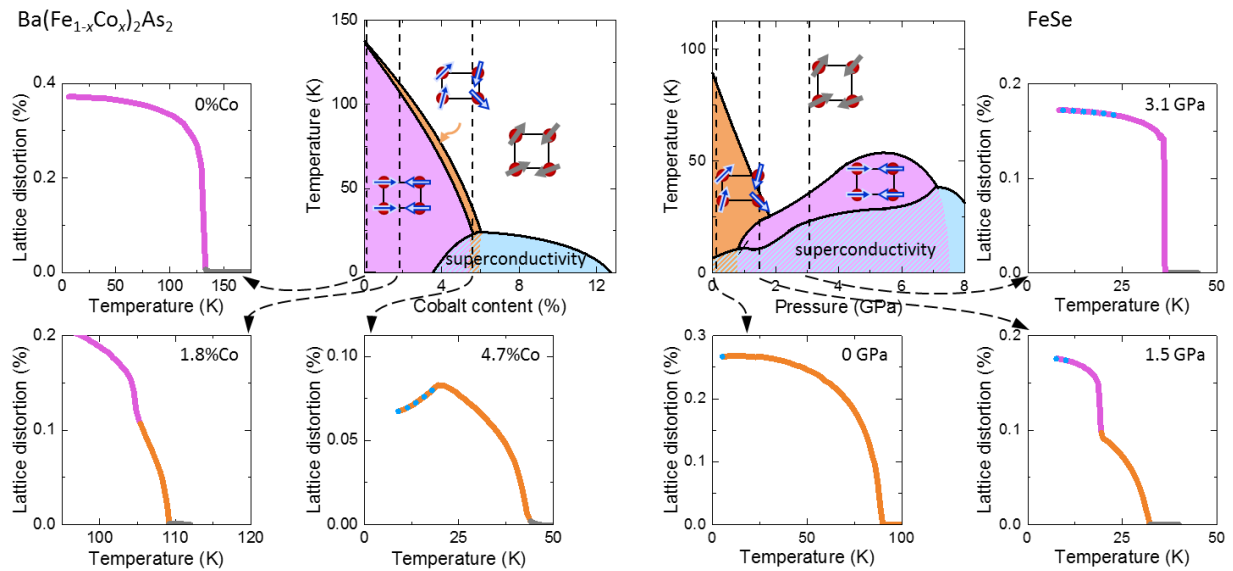

Fig. 5. Lattice distortion (relative length difference between the two directions in a rectangular lattice) in cobalt-doped BaFe$_2$As$_2$ and in FeSe under pressure, measured along the dashed lines in the respective phase diagram. Colors correspond to specific phases.

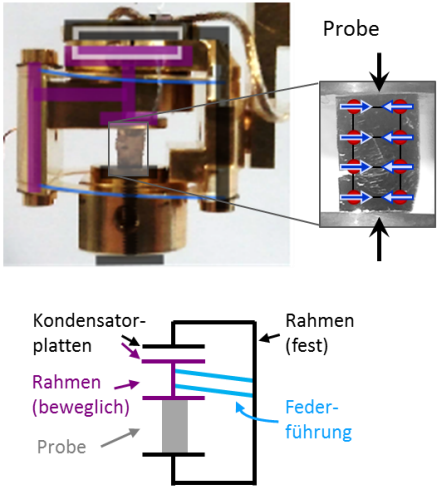
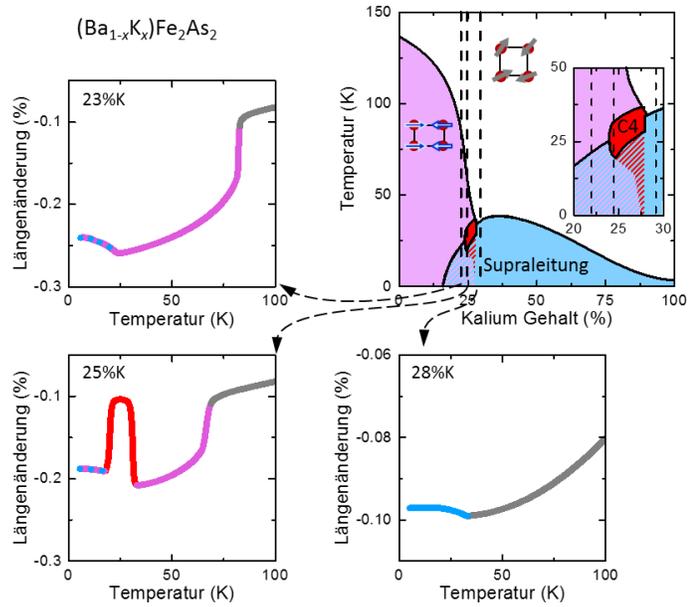

Abb. 6. (Links) Foto und Aufbau eines kapazitiven Dilatometers zur hochaufgelösten Messung von Längenänderungen. (Rechts) Damit gemessene Längenänderungen von Kalium-dotiertem BaFe$_2$As$_2$ entlang der gestrichelten Linien im Phasendiagramm, mit besonderem Augenmerk auf die rot makierte C4 Phase. Fotos: Kai Grube, Liran Wang

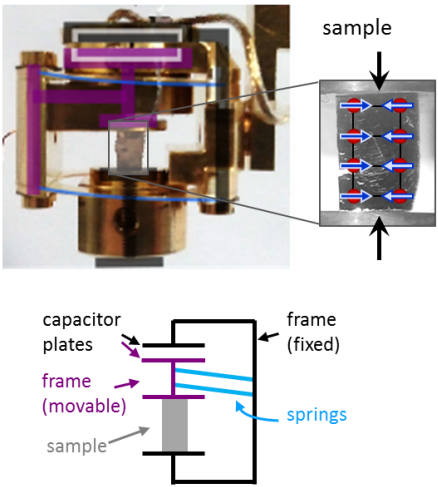
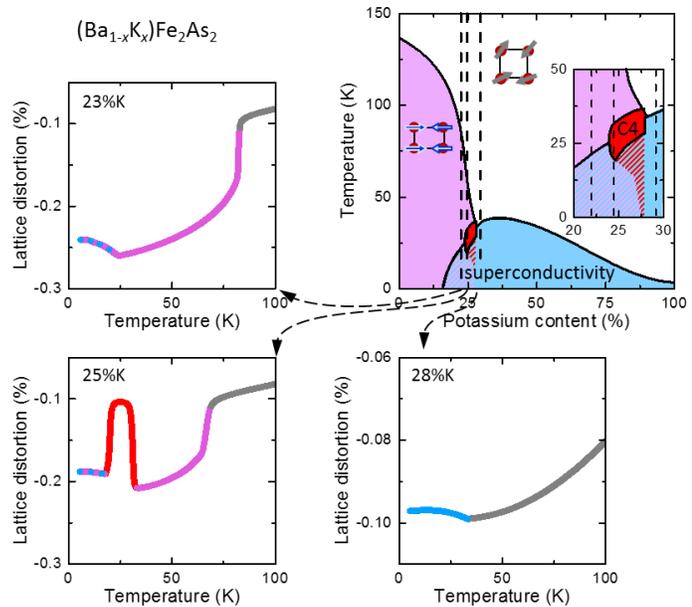

Fig. 6: (left) Photograph and schematic illustration of a capacitance dilatometer for the high-resolution measurement of length changes. (right) Length changes of potassium-doped BaFe$_2$As$_2$ measured with the dilatometer along the dashed lines in the phase diagram. The C4-phase is marked in red.
Photos: Kai Grube, Liran Wang